# Further improvement of lattice thermal conductivity from bulk crystalline to 1-D-chain polyethylene: A high-yet-finite thermal conductivity using first-principles calculation

Xinjiang Wang[a], Massoud Kaviany[b] and Baoling Huang[a,*]

We calculate the thermal conductivity ($\kappa$) of both bulk crystalline and single-chain polyethylene (PE) using the first-principles-based anharmonic lattice dynamics. Despite its low $\kappa$ in amorphous state, the predicted bulk crystal has high axial $\kappa$ (237 W/m-K) at room temperature. The much lower measured $\kappa$ is attributed to the small size of nanocrystallites (~10 nm) in synthesized semi-crystalline PE. For the 1-D chain, the predicted $\kappa$ is much larger and yet finite (1400 W/m-K at room temperature). The reduction of scattering phase space caused by the diminished interchain van der Waals interactions explains this larger $\kappa$. It is also found that the transverse phonon branches with quadratic dispersion make minor contribution to this, due to their vanishing group velocity in the long-wavelength limit. Moreover, the low-frequency bending and twisting phonon modes are strongly coupled and dominate anharmonic phonon scatterings, leading to the finite $\kappa$. The predicted high $\kappa$ of bulk and chain PE crystals enable polymer usage in thermal management and the above phonon scatterings provide guide for their nano-designs.

## Introduction

Polymers are classified as thermally resistive due to low thermal conductivity (0.1 to 0.5 W/m-K) in amorphous state[1]. However, in applications ranging from cell-phone cover to encapsulation of solar cells[2], effective heat dissipation is imperative (overheating destabilizes device performance and reduces lifetime). So, there is interest in thermally conductive polymers, and polymers with thermal conductivity $\kappa$ above 10 W/m-K are already competitive as *in-situ* heat sink in LED devices[3], and polymers can be corrosion resistant, adaptive to harsh conditions as in battery cells[4]. Most improvements to $\kappa$ of polymers have been with composites of high-$\kappa$ nanoparticles, such as BN in epoxy reaching $\kappa$ as high as 10 W/m-K[1], yet still far below prediction from the mixture model. This upper limit is controlled by the particle-particle contact resistance (which is much larger for van der Waals contacts than any formed even with weakest covalent bonds[5,6]). Another improvement is crystallization by mechanical stretching[7], electrospinning[8] or molecular layer deposition[9], and increase of axial thermal conductivity of polymer with large draw ratio have long been observed[10] ($\kappa \approx 42$ W/m-K at room temperature with draw ratio ~ 350[11]). This is because the structure of polymers, for example polyethylene (PE), becomes a good conductor in its axial direction due to strong carbon-carbon covalent bonds (approaching C-C bonds in diamond and graphite[12]). Since experiments with ideal polymer crystallization remain challenging, molecular dynamic (MD) simulations have been used and lattice thermal conductivity of PE (amorphous[13,14] and crystalline[14–16]) has been predicted with axial $\kappa$ of the crystalline structure from 47[15] to 310 ± 190 W/m-K[16], at the room temperature.

Improvement in thermal conductivity has also been found through low dimensionality, including delamination of single-atomic layers from stacked bulk crystals, e.g., single-layer graphene and h-BN have larger thermal conductivity than their bulk lattices[12,17]. This is due to reduction of scattering channels (symmetry constraints[18] and removal of inter-layer coupling governed by strongly anharmonic van der Waals forces[19]). So there has been interest in thermal transport in single polymer chains, and thermal conductivity reaching 104 W/m-K was recorded for PE nanofibers of diameter around 50 nm[7]. Thermal conductance of chains containing tens of units of alkane thiols has also been successfully measured using scanning thermal microscopy[20] and femtosecond laser pulse[21], and interest continues in single chain extraction[22] and nanoscale thermal conductivity measurement[23]. The quasi-1-D lattice (containing transverse motion) $\kappa$ has been predicted by atomic simulations, e.g., MD[14,15,24–28] and the Green function method[29], with larger values compared to bulk lattice, while bifurcation over the convergence of an infinite chain has been reported. The $\kappa$ of PE chain increases with length within the maximum simulation size (1000 unitcell), while a converged thermal conductivity is found for poly(p-phenylene)[27]. A polydimethylsiloxane (PDMS) chain is even reported to possess a low thermal conductivity of 7 W/m-K[14]. The equilibrium MD results for 1-D PE chains are also not conclusive: convergent $\kappa$ is reached ~3 ns[26] for some cases, while divergence is found under different initial conditions[28]. This debate actually predates the investigations on polymer chains, whether thermal conductivity of the 1-D lattices converges[30–32] or not[33,34].

Previous computational studies predicted $\kappa$ and unveiled transport mechanisms. However, the Green function method[29] assumes ballistic phonons transport except at the boundaries and predicts diverging $\kappa$ with increasing simulation size (thus applicable at low temperatures or small system size). On the other hand, the classical MD simulation uses problematic empirical potentials, e.g. different treatments of hydrogen atoms in polymer[35], and since the axial group velocity of polymer crystal is large[36] (leading to a Debye temperature > 1000 K) the room-temperature MD simulations are questionable. The non-equilibrium MD simulations require simulation dimensions much larger than the largest phonon mean free path (MFP)[37], while equilibrium MD requires sufficiently large simulation time, which is challenging under small intrinsic scattering strength[38].

[a.] Department of Mechanical and Aerospace Engineering, The Hong Kong University of Science and Technology, Clear Water Bay, Kowloon, Hong Kong. Email address: mebhuang@ust.hk
[b.] Department of Mechanical Engineering, University of Michigan, Ann Arbor, MI 48109, USA

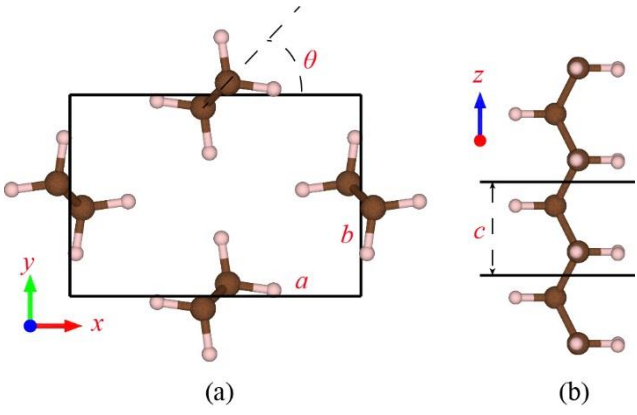

Figure 1. (a) The structure of bulk PE crystal in the cross-axial plane, and (b) structure of a single-chain (1-D) PE crystal.

Here we use the first-principles calculations and solve the Boltzmann transport equation (BTE) to find the thermal conductivity of bulk crystalline polyethylene (PE) and the PE chain and the convergence property. The fully relaxed PE crystal have an orthorhombic structure as the one shown in Fig. 1. The parameter $a$, $b$ and $c$ are the lattice constants and $\theta$ is the chain setting angle (the dihedral angle between the plane of carbon atoms on a single chain with the $xz$ plane).

This technique has been successfully used for bulk [39,40] and low-dimensional materials [38,41,42] and has the advantage of accuracy and no fitting parameters (solving the electron Kohn-Sham equation [43]). The BTE enables spectral contribution to the thermal conductivity and expression of scattering mechanisms directly [44,45].

## METHODS

**Lattice thermal conductivity**

Calculation of lattice $\kappa$ is reviewed elsewhere [46–49], and interatomic force constants are the most important input to determining the intrinsic phonon properties and scattering mechanisms. Upon expanding the total lattice potential energy $E$ to the third order, the harmonic ($\Phi$) and cubic anharmonic force constants ($\Psi$) are derivatives of $E$ with atomic displacements $U$

$$\Phi^{ij}_{lb,l'b'} = \frac{\partial^2 E}{\partial U^i_{lb} \partial U^j_{l'b'}} , \qquad (1)$$

$$\Psi^{ijk}_{lb,l'b',l''b''} = \frac{\partial^3 E}{\partial U^i_{lb} \partial U^j_{l'b'} \partial U^k_{l''b''}} , \qquad (2)$$

where $l$ and $b$ are the indices of supercell and the atom inside a unitcell and $i$, $j$, $k$ are Cartesian indices. Both $\Phi$ and $\Psi$ can be calculated under the finite displacement method by replacing the partial derivative operators with small finite displacement in Eqs. (1) and (2) and $E$ can be accurately predicted using first-principles method.

Under the harmonic approximation, the lattice vibrational dynamical matrix is obtained from the harmonic interatomic force constants

$$D^{ij}_{bb'}(\mathbf{q}) = \frac{1}{N\sqrt{m_b m_{b'}}} \sum_l \Phi^{ij}_{0b,l'b'} e^{i\mathbf{q}\cdot R^{l'b'}_{0b}} , \qquad (3)$$

where $\mathbf{q}$ is the wavevector, $R$ is distance between two atoms. Then phonon frequency $\omega$ and eigenvector $\mathbf{e}$ are obtained by diagonalizing the dynamical matrix

$D(q)\mathbf{e}(qp) = \omega^2(qp)\mathbf{e}(qp)$, where $p$ is the phonon polarization index and each wavevector $q$ and polarization $p$ define a phonon $\lambda$. The equilibrium phonon Bose-Einstein occupancy is $n^o_\lambda = \left[1 - \exp(-\hbar\omega/(k_B T))\right]^{-1}$. Heat flows due to deviation from equilibrium, $n_\lambda - n^o_\lambda = n^o_\lambda(n^o_\lambda+)\mathbf{F}(qs)\cdot\nabla T$, where $\nabla T$ is the temperature gradient and $\mathbf{F}$ is the linear phonon perturbation vector. Then thermal conductivity (Fourier law) is sum of contributions from all phonons, i.e.,

$$\kappa_{ij} = -\frac{1}{N_o \Omega} \sum_\lambda \hbar \omega_\lambda v^i_\lambda n^o_\lambda (n^o_\lambda + 1) F^j_\lambda , \qquad (4)$$

where $v$ is the phonon group velocity $v^i_{qs} = \partial\omega(qs)/\partial_i q$, $\Omega$ is unitcell volume and $N_o$ is total number of $q$-points (within first Brillouin zone), and $\mathbf{F}$ can be obtained solving the BTE

$$-v^i(\lambda)\frac{\partial n^0_\lambda}{\partial T} = \frac{1}{2}\sum_{\lambda',\lambda''}\left[\left(P^{\lambda''}_{\lambda,\lambda'} + P^{\lambda'}_{\lambda,\lambda''} + P^\lambda_{\lambda',\lambda''}\right)\left(F^\alpha_\lambda + F^\alpha_{\lambda'} + F^\alpha_{\lambda''}\right)\right] + P^b_\lambda F^i_\lambda , \qquad (5)$$

where $P^{\lambda''}_{\lambda,\lambda'}$, $P^{\lambda'}_{\lambda,\lambda''}$ and $P^\lambda_{\lambda',\lambda''}$ are three-phonon scattering and $P^b_\lambda$ boundary scattering probability. All three-phonon scattering processes satisfying $\mathbf{q} + \mathbf{q}' + \mathbf{q}'' = \mathbf{0}$ ($\mathbf{G}$), where $\mathbf{G}$ is the reciprocal lattice vector. The above expression is slightly different from some literature [47,50] but consistent with Chaput's simplification [51], where in principle this definition is equivalent considering the time reverse symmetry while this definition makes the expression of Eq. (5) more elegant and the implementation easier [51]. Without loss of generalization, only $P^{\lambda''}_{\lambda,\lambda'}$ is written as

$$P^{\lambda''}_{\lambda,\lambda'} = 2\pi n^o_\lambda n^o_{\lambda'} (n^o_{\lambda''} + 1) \left|V_{\lambda,\lambda',\lambda''}\right|^2 \delta(\omega_\lambda + \omega_{\lambda'} - \omega_{\lambda''}) , \qquad (6)$$

where $\delta$ is the Dirac delta function and $V$ is the anharmonic force constants projected onto the eigenvector space, i.e.,

$$V_{\lambda,\lambda',\lambda''} = \left(\frac{\hbar}{8N_0\omega_\lambda\omega_{\lambda'}\omega_{\lambda''}}\right)^{1/2} \cdot \sum_{\substack{bb'b''\\ijk}} \sum_{l'l''} \Psi^{ijk}_{0b,l'b',l''b''} e^{iq'\cdot R^{l'b'}_{0b}} e^{iq''\cdot R^{l''b''}_{0b}} \frac{\mathbf{e}^i_{\lambda b}\mathbf{e}^j_{\lambda' b'}\mathbf{e}^k_{\lambda'' b''}}{\sqrt{m_b m_{b'} m_{b''}}} . \qquad (7)$$

For the case of finite system, phonons experience an additional scattering with the boundary. Assuming a diffuse surface there with elastic scattering, we have [46]

$$P^b_\lambda = \frac{v_\lambda}{L} n^o_\lambda (n^o_{\lambda'} + 1) , \qquad (8)$$

where $L$ is the Casimir effective length of the sample.

With the scattering operators, Eq. (5) is solved first with neglecting $\mathbf{F}_{\lambda'}$, $\mathbf{F}_{\lambda''}$ and the off-diagonal components of the scattering matrix [single-mode relaxation-time model (SMRT) [52]] and later iteratively to obtain the exact solution using preconditioned conjugate-gradient method [49] for its less stringent and faster convergence.

With the full solution of BTE, we can define an effective phonon lifetime

$$\tau_{eff,\lambda} = \frac{k_B T^2}{\hbar \omega_\lambda} \cdot \frac{\mathbf{F}_\lambda \cdot \mathbf{v}_\lambda}{|v_\lambda|^2} \qquad (9)$$

by comparing Eq. (4) with the general thermal expression from SMRT $\kappa_{ij} = \frac{1}{N_o \Omega} \sum_\lambda C_v(\lambda) v_\lambda^i v_\lambda^j \tau_\lambda$, where $C_v$ is modal heat capacity. Different from the relaxation time from SMRT, this newly-defined lifetime conveys information of multiple phonon excitations and yet casts a simplified picture in the conventional frame for better understanding.

**Quasi-harmonic approximation**

In order to verify accuracy of the potentials used in anharmonic properties, the predicted thermal expansion ($\alpha$) of PE crystal is compared with the experiments, since $\alpha$ arises from the anharmonic lattice vibration. This is realized by minimizing the Helmholtz free energy with respect to lattice parameter. Under quasi-harmonic approximation (linear relation of phonon frequency change and lattice constants), we have

$$\alpha_i = \frac{1}{N_0 \Omega} \sum_\lambda C_v(\lambda) S_{ij} \gamma_j(\lambda), \qquad (10)$$

where $S$ is compliance tensor, and $\gamma_i = \partial \ln \omega_\lambda / \partial \ln a_i$ is the diagonal components of modal Gruneisen tensor.

**First-principles calculations**

For bulk PE, both relaxation and static energy calculations use the van der Waals (vdW) functional optB88-vdW[53,54] in addition to the local-density approximation, as implemented in the *Vienna ab initio simulation package* (VASP) [55,56]. The structure was fully relaxed to equilibrium position with a cutoff energy of 550 eV and 11×15×31 **k**-point meshing scheme in the Brillouin zone. Then the original cell was expanded to a supercell of the size 2×3×5 for the harmonic force constants calculation and 2×2×3 for the anharmonic ones. Interaction cutoff distances were used during the anharmonic force constants calculation of crystalline PE to reduce the computation without impairing accuracy. An interaction distance of 8 Å was selected for C atoms while that for H atoms was 6 Å. The thermal conductivity was calculated on different mesh sizes until convergence reached. The Dirac delta function in Eqs. (6) and was evaluated based on the linearized tetrahedron method[48,57]. This method circumvents the artificial smearing factor used in the smearing method, which approximates the Dirac delta function as a Gaussian distribution[47].

For PE chain, the first-principles calculations is similar to the bulk, except a vacuum slab of thickness 17 Å, large enough to diminish the influence of vdW forces, was used to accommodate the chain in both *a* and *b* directions. The supercell size for harmonic and anharmonic properties were chosen as 1×1×9 and 1×1×5, respectively. Although the calculation process on 1-D PE chain should be simpler than the bulk (because of the disregard for dimensions *a* and *b*), the results are more sensitive to the inevitable errors during the calculation. So special techniques were used for calculations of PE chain.

**Symmetry constraints on interatomic force constants**

The symmetries constraints on both harmonic and anharmonic force constants include permutational, translational, rotational invariances and point group symmetries[58]. These symmetry constraints on the system not only reduce the number of calculations, but allow verifying the physicality of results. There are intrinsic errors in the first-principles calculations, such as incomplete basis wavefunctions and small-but-finite residual forces in the relaxed structure. These small errors influence the behavior of long-wavelength phonons significantly (i.e., dispersion in long-wavelength limit). Therefore, the calculated force constants need to be adjusted to be physically reasonable, but such adjustments should be minute to keep the major lattice properties. There have been multiple methods to symmetrize the force constants, including the Lagrange multiplier method[59,60] and internal coordinate method[61]. Here we use a least-squares method which is similar to the Lagrange multiplier method but easier to implement. First, the force constants $P$ (either $\Phi$ or $\Psi$) were reduced under the symmetry invariances to $N$ independent elements $\theta_j$ (with $j$ ranging from 1 to $N$) . The force constants is retrieved from $\theta_j$ through

$$P_i = \sum_j A_{ij} \theta_j \qquad (11)$$

where $A$ is the transformation matrix and $i$ is the index of the original force constants. By utilizing the least-squares method, which implies least discrepancy of the newly obtained force constants from the original, the irreducible elements were solved and the force constants satisfying all the symmetry constraints were deduced using Eq. (11).

**Dirac delta function for 1-D mesh**

The solution to BTE, Eq. (5), involves integrations of scattering rates in the Brillouin zone and the Dirac delta function estimation in Eq. (6). This is conventionally accomplished by assuming a Gaussian distribution of the Dirac delta function with an arbitrarily assigned smearing factor, which is either constant[49] or adaptive[47]. But this method is accurate only when the mesh size approaches infinite and the smearing factor approaches zero, thus requiring a bilateral test of both mesh and smearing factor. Apart from the demanding computations, the results are found sensitive to the choice of the smearing factor under high thermal conductivity, so we use the linear tetrahedron method[48,57,62] for the Dirac delta function. Therefore, the calculations are free from the smearing factor and the accuracy of the results depends only on the mesh size, and this method adapts naturally to 1-D mesh in the form of $1 \times 1 \times N_o$. Starting from Refs. [57,62], in 1-D case, the estimation of $\delta(\omega - \omega_q)$ is

$$\delta(\omega - \omega_q) = \sum_{j=1}^{2} g_{q,j}(\omega) , \qquad (12)$$

where $g$ is evaluated on the $q$-point together with its two neighbors indexed by $j$.

$$g_{q,j} = \begin{cases} \dfrac{|\omega - \omega_j|}{(\omega_q - \omega_j)^2} & \text{if } \min(\omega_j, \omega_q) < \omega < \max(\omega_j, \omega_q) \\ 0 & \text{else} \end{cases} \qquad (13)$$

Difficulty with this method become more serious with a 1-D mesh. First, energy conservation may be satisfied intrinsically [i.e., the denominator $\omega_q - \omega_j = 0$ in Eq. (15)], as illustrated by a collinear scattering events in the three-phonon processes on a linear dispersion relation (e.g., longitudinal acoustic). Then the estimated delta function for 1-D mesh from Eqs. (12) and (13) would be infinite, while in a 3-D mesh the chance that the four corners of a tetrahedron have the same frequency is much slimmer. Another problem with the above estimation is its only dependence on the gradient of dispersion relation at *q*-point. Therefore, if there is an exchange of integration sequence of $q$ and $q'$ in $\iint F_{q,q'}\delta(\omega_q - \omega_{q'})dqdq'$, the estimation using tetrahedron method leads to $\delta(\omega_q - \omega_{q'}) \neq \delta(\omega_{q'} - \omega_q)$. Especially, prominent in three-phonon scatterings, the asymmetric delta function breaks the interchangeability of the three phonons in the scattering probability $P_{\lambda,\lambda'}^{\lambda''}$. This would further lead to the breakdown of the positive definite property of the collision matrix[51], which could possibly result in a false divergent thermal conductivity calculation.

This is solved by realizing the double integration and building a hyperspace $q \otimes q'$, which is discretized subsequently in this higher dimensional space ($\iint F_{q,q'}\delta(\omega_q - \omega_{q'})dqdq' \rightarrow \int_{q \otimes q'} F_{q,q'}\delta(\omega_q - \omega_{q'})dS_{q,q'}$). This process theoretically turns the tetrahedron for single integration in 3-D space into a 6-symplex (a counterpart of tetrahedron in 6-D space). Then the 6-symplex enables interchanging the integration sequence for any two dimensions and thus the Dirac delta function is symmetric again. Evaluation of the delta function would be tedious in 6-D space, and fortunately the breakdown of asymmetry is not so significant for a 3-D bulk material. For the 1-D PE chain however, the hyperspace is only a 2-D space on which the Dirac delta function is readily and elegantly evaluated.

## RESULTS AND DISCUSSION

### Bulk PE crystal

The predicted lattice parameters are listed and compared with experiments at different temperatures in Table I, with agreement within 2%, verifying the first-principles calculations for the bulk PE crystal.

Table I. Calculated lattice parameters for the bulk PE compared with experiments at few temperatures.

| Lattice parameters | Predicted | Experiment (4 K)[63] | Experiment (10 K)[64] | Experiment (77 K)[65] |
|---|---|---|---|---|
| $a$ (Å) | 6.978 | 7.121 | 7.16 | 7.155 |
| $b$ (Å) | 4.854 | 4.851 | 4.86 | 4.899 |
| $c$ (Å) | 2.553 | 2.548 | 2.534 | 2.547 |
| $\theta$ (°) | 43.3 | 41 ± 1 | | |

Starting from the equilibrium positions, the phonon dispersion and density of states (DOS, $D_p$) of the PE crystal are shown in Figs. 2 (a) and (b), up to cutoff frequency of 90 THz, but most phonons are localized (small group velocity, exemplified by phonons with wavevector perpendicular to the chain axis and frequencies larger than 40 THz). This small group velocity is from the C-H bonds and the localized phonon across the chain direction by the weak vdW interactions between the neighboring chains. The group velocity of acoustic phonons along the chain is rather high (17 km/s, close to that of diamond and carbon nanotube), due to the strong axial C-C covalent bonds, which also accounts for the large axial Young modulus (330.7 GPa). So, $\kappa$ across the chain axis and the contribution from high-frequency phonons may be negligible. The dispersion along Γ-Z with a clearer insight is shown along with the experiments in Fig. 2(c), in good agreement (validates the predicted harmonic interatomic force constants). The eight branches shown in Fig. 2(c) are the split vibrational modes from the 4 acoustic branches of a single PE chain (to be discussed later).

The eigenvectors of phonon modes belong to a specific group and can be distinguished in Fig. 2(c) with different colors (Detailed assignment of phonon modes of the PE crystals are also found in litertarure[66]). The phonon modes cross linked with low group velocity and frequency lower than 5 THz are the torsional and transverse modes (small twisting and bending rigidity of the PE chain). The phonon dispersion crossing enlarges space of the three-phonon scattering (leading to a small lifetime and small contribution to $\kappa$).

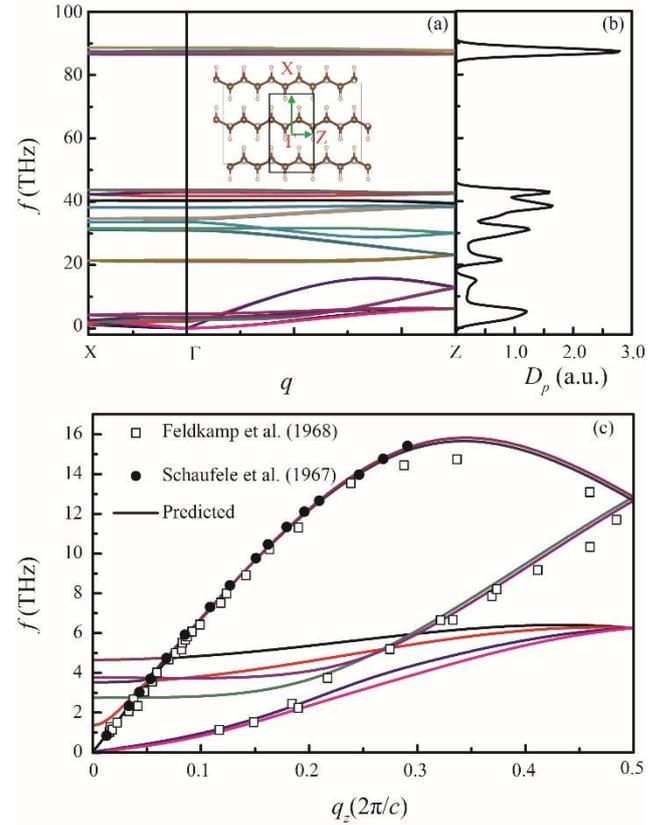

Figure 2. (a) Predicted phonon dispersion of bulk PE crystal along different directions marked in the insets, (b) phonon density of states ($D_p$), and (c) comparison of predicted dispersion along Γ-Z with the experiments[67,68]. The Feldkamp et al.[67] experiments have been rescaled from deuterated PE crystal. The colors denote phonon branches along a specific direction.

The anisotropic thermal expansion $\alpha$ is important and derives from the lattice anharmonicity and with high crystallization ratio $\alpha$ varies little with increase in crystallinity at low temperatures[69]. Therefore, despite the lack of $\kappa$ data, comparisons of predicted $\alpha$ with low temperature experiments can help validate the anharmonic potentials used in $\kappa$. Figures 3(a) to (d) show predicted directional $\alpha$ compared with X-ray experiments[70–72] for temperatures up to 200 K [higher than the sub-glass transition temperature of high-density polyethylene(~145 K)[69]]. The change of directional $\alpha$ with temperature is well captured, especially along the *ab* plane. The difference in the axial direction [Fig. 3(c)] is rather small at temperatures lower than 100 K (but increases at higher temperatures). The predicted axial $\alpha_c$ is negative at temperatures lower than 200 K (tends to saturate and slightly increase for $T > 150$ K), but measured value decrease over the temperature range, and can be due to the imperfect crystallization of the samples (~70% in Ref [72]). The taut-tie molecules among microcrystals resemble fully crystalline PE in thermal expansion behavior at low temperatures, but lead to contraction of the separation distance of crystallites when tie molecules vibrate with large amplitude (resulting in large negative thermal expansion at temperatures higher than the sub-glass transition point). For ideal crystal, this secondary effect caused by the tie molecules does not exist (predicted axial thermal expansion is higher than the experiments at around 200 K), but the agreement in general is good.

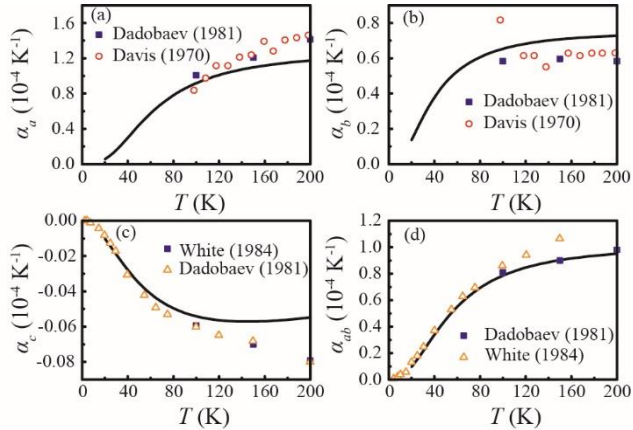

Figure 3. The variations of predicted anisotropic thermal expansion coefficient with temperatures (from quasi-harmonic approximation) along axis (a) *a*, (b) *b*, and (c) *c*, and (d) plane *ab*, and compared with experiments[70–72].

Figure 4(a) shows the predicted directional $\kappa$ as a function of temperature, in all directions. Both axial and perpendicular $\kappa$ have been mesh-size independence verified. The results from the exact solution of BTE are only larger than the SMRT model prediction by around 20% for $T > 200$ K, but for $T < 200$ K the difference is large. This is because high-frequency phonon modes, including those at the edge of Brillouin zone, are not excited at $T < 200$ K, which significantly reduces the Umklapp 3-phonon scatterings. Therefore the normal processes dominate the scattering events and lead to poor prediction by the SMRT. The predicted axial $\kappa_{zz}$ is 237 W/m-K at room temperature, in between the classical MD prediction values of 47 to 310 W/m-K[15,16] but significantly higher than all recorded experimental values (e.g. 104 W/m-K[7], 41 W/m-K[11] and 27 W/m-K[73]). As with the thermal expansion, this can be due to the non-ideal crystal in experiment. The axial crystallite length ($l_z$) of semi-crystalline PE is rather small (8-42 nm[71]). Therefore, the crystallite boundaries suppress the phonon transport and lead to a much smaller thermal conductivity. Once the phonon boundary scatterings from the crystallite boundaries are included with the same length (20 nm) measured in Ref. [73] through Eq (8), the calculated $\kappa_{zz,b}$ (axial thermal conductivity with boundary scatterings) well predicts the measurements in the same study. Since the size of axial crystallites is a function of draw ratio and preparation method and may reach as high as 3 μm[11], experimentally recorded $\kappa_{zz}$ results spans rather widely as shown in Fig. 4(a). The coincidence between other predicted $\kappa_{zz,b}$ with fitted crystallite lengths and measurements on different PE samples reveals different estimated sample qualities and the potential of further development.

It is also observed that $\kappa$ is strongly anisotropic. In Fig. 4(b), the cross-axial value is almost three orders lower than the axial value. Small $\kappa_{xx}$ and $\kappa_{yy}$ are due to the weak vdW bonds across chains (low group velocity and strong anharmonic scattering). The average cross-axial predicted $\kappa_{xx}$ and $\kappa_{yy}$ (i.e., the normal thermal conductivity $\kappa_n$) is 0.45 W/m-K at room temperature, twice as much as that from experiments (~0.22 W/m-K[36,74]). This discrepancy becomes more pronounced at lower temperatures, which is also attributed to the small thickness of the nanocrystallites (~13 nm[73]) of semi-crystalline PE structure in experiments. The inclusion of the boundary effect with the above thickness retrieves the trend of measured data at different temperatures. Unlike $\kappa_{zz}$, experimental $\kappa_n$ results are quite consistent from different studies, which might be due to the smaller variance of cross-axial crystallite thickness compared with the axial length even with different draw ratios.

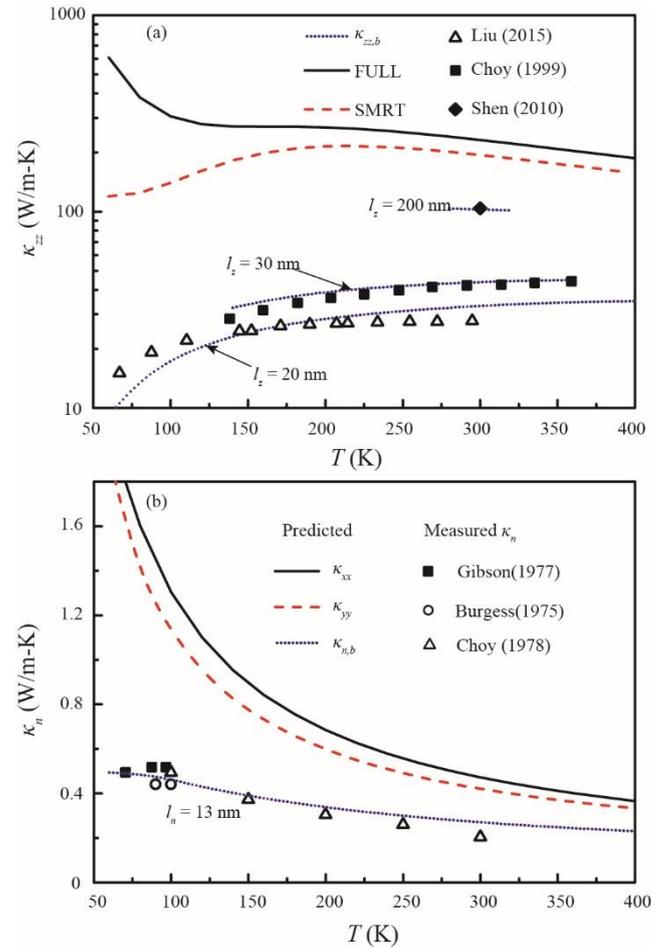

Figure 4. (a) Variations of axial thermal conductivity of PE bulk crystal with temperature. The full BTE and the single-mode relaxation-time (SMRT) model results are shown. Axial thermal conductivity accounting for boundary scatterings ($\kappa_{zz,b}$) with different lamina lengths ($l_z$) are compared with previous measurements[7,11,73]. $l_z = 20$ nm is given in the compared experimental study[73] while $l_z = 30$ nm and $l_z = 200$ nm are fitted from the corresponding experimental studies in comparison, respectively. (b) Same as (a) for cross-axial thermal conductivity ($\kappa_{xx}$, $\kappa_{yy}$ and $\kappa_n$), where $\kappa_n$ is normal thermal conductivity to the chain axis. Predicted normal thermal conductivity of nano crystallite ($\kappa_{n,b}$) adopts the crystallite thickness ($l_n$) value of 13 nm also given by Ref. [73]. Results from SMRT vary little from the full BTE solution for cross-axial thermal conductivity and are thus not shown. Dotted points are $\kappa_n$ values extracted from experimental studies [10,77,78].

**PE chain crystal**

As the bulk PE crystal is reduced to a single chain, i.e., the molecules are free from the vdW interactions with neighboring chains, the high-frequency vibrational modes barely change, but the low-frequency dispersions are affected, as shown in Fig. 5. For example, the cutoff frequency of the torsional mode in 1-D PE shifts down by around 1 THz due to the depletion of vdW constraint in the cross-axial direction. Besides, there are multiple optical phonon modes in the bulk PE in the low-frequency range, but dispersion reduces to 4 acoustic modes in a single chain, i.e., two translational acoustic modes (TA1 and TA2), a torsional acoustic mode (TWA), and a longitudinal acoustic mode (LA).

The lack of split optical modes in 1-D PE has multiple consequences. The optical modes in bulk PE increase the phonon DOS, while decreasing the effective phonon group velocity [$f$ = 2~5 THz in Fig. 5(b)] at the corresponding frequencies. The multiple intertwined optical modes also broaden the phase space of the acoustic phonon scatterings, which is more prominent when all dispersion curve from ($x, y, 0$) to ($x, y, \pi/c$) ($x$ and $y$ are arbitrary) in the Brillouin zone of bulk PE are projected and shown as the shaded area. A spectral analysis on the vibrational eigenmodes reveals that the flat optical modes are mainly the hybridized twisting mode with transverse modes. The hybridization of twisting mode with other acoustic modes has also been reported to significantly increase the phonon scattering in quasi-1-D structures[77]. It is expected that phonon transport in the bulk PE is similarly compromised.

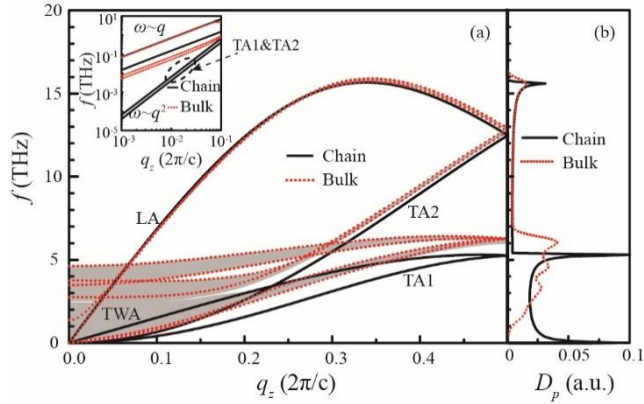

Figure 5. (a) Phonon dispersion of 1-D PE chain (along the chain) at low frequencies. TA1 and TA2 are the two transverse phonon modes vibrating perpendicular to the chain axis, TWA is the twisting and LA is the longitudinal mode. The phonon dispersion of the bulk PE crystal from Γ to Z is shown as dotted curves while all dispersion curves of the bulk PE from $q_z=0$ to $q_z=\pi/c$ are projected as the shaded area. The inset compares the dispersion at the vicinity of Γ in logarithmic scale. (b) Phonon DOS of 1-D chain and bulk PE. (DOS is normalized to make the total number of states equal to one).

The difference in dispersion is more influential in the transverse modes at long wavelengths, as shown in inset of Fig. 5(a). The two TA modes of 1-D PE chain have quadratic dispersion, while the torsional and longitudinal modes are linear. In comparison, the dispersion of bulk PE is linear for acoustic phonons at small wavevectors. For a fully relaxed 1-D structure without any internal stress, the 1-D chain is an infinitely long rod in the long-wavelength limit and acoustic vibrations degrade to the classic continuum model for rods with quadratic transverse dispersions[78]. Therefore, the dispersion relations of the two TA modes are expected to be quadratic in 1-D chain structure, the same as the ZA mode in 2-D structures such as graphene and borophene[61]. This difference impacts the DOS ($D_p$) at low frequencies. From the definition of $D_p$ in Ref [46], the bulk PE with linear dispersion relation follows a relation of $D_p \sim \omega^2$, while for 1-D chain with a quadratic dispersion the relation is $D_p \sim \omega^{-1/2}$. This difference is illustrated in Fig. 5(b), where the $D_p$ of bulk PE converges to 0 for frequencies lower than 1 THz, while for the 1-D PE it tends to diverge with decrease in frequency.

In spite of the diverging $D_p$ of the 1-D PE chain, the contribution of these phonons to $\kappa$ is not ascertained to diverge. Using the kinetic theory $\kappa = \int D_p(\omega) C_v(\omega) v(\omega) \Lambda(\omega) d\omega$, where $\Lambda$ is the phonon mean free path, even though $D_p$ diverges as $\omega^{-1/2}$, $v \sim \omega^{1/2}$ at low frequencies so they cancel and divergence depends on the behavior of $\Lambda$. Since the anharmonic properties are calculated from the anharmonic interatomic force constants, $\kappa$ is obtained and shown in Figs. 6(a) and (b). The small variation of $\kappa$ with mesh size, as shown in the inset of Fig. 6(a), validates the applicability of tetrahedron method in 1-D chain and the convergence with discretization in the Brillouin zone. It is seen that the SMRT model appears a poor estimate for the thermal conductivity at all temperatures for its underestimation of $\kappa$ by at least 50% (even >99 % at low temperatures). Such underestimation is also a result of dominant normal (momentum-conserving) phonon scattering processes as in the bulk PE at low temperatures and other 2-D materials[79] since simple treatment of normal process as a dissipation source significantly underestimate the thermal conductivity. The large $\kappa$ of an infinite chain at different temperatures are still finite and the predicted value at room temperature is 1400 W/m-K. Even though the absolute value of $\kappa$ does not reach other carbon-based materials, e.g., graphene and carbon nanotubes[12], it is more than 3 orders of magnitude higher than that of amorphous PE and 6 times the bulk PE, mainly caused by the reduction of scattering phase space due to the disappearance of inter-chain vdW forces.

The $\kappa(T)$ peaks around 145 K, which is due to the counterbalance between the heat capacity and the anharmonic scattering, especially for the LA and TA2 phonon modes. At high temperatures, all the phonons are excited and the heat capacity is saturated and $\kappa$ decreases with temperature (stronger anharmonic scattering), while at lower temperatures the heat capacity decreases (reduction of excited phonons impairs transport). Especially, LA and TA2 phonons with higher cutoff frequency are more sensitive to the decrease of temperature. It is also noted that the thermal conductivity comes mainly from LA phonons at high temperatures until a crossover between LA and TA2 at T<100 K. It is because of the large group velocity of LA phonons and low scattering rate at high frequencies, which would be revisited in Fig. 7.

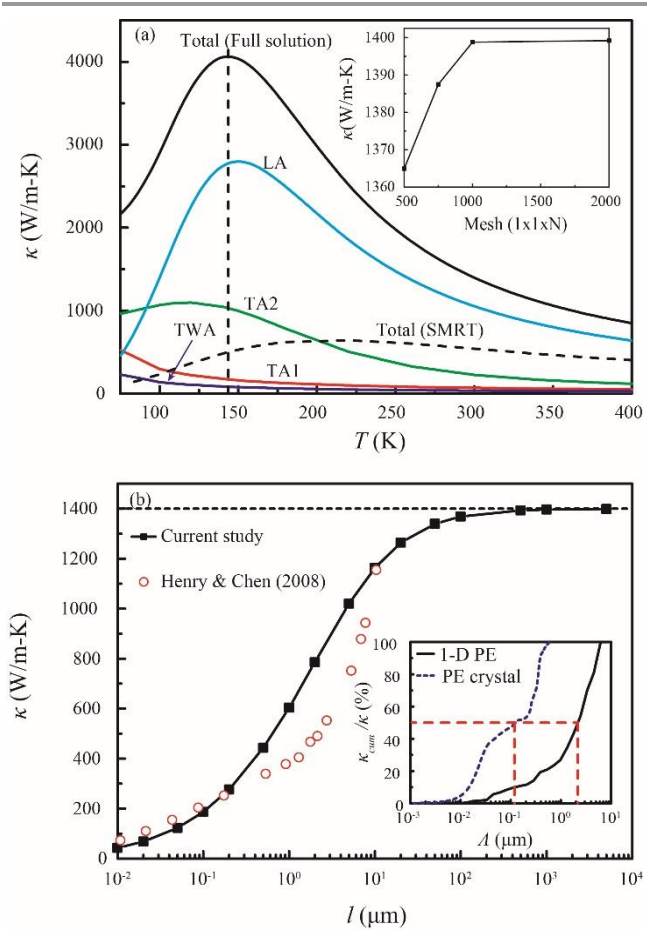

Figure 6. (a) Variations of the axial thermal conductivity of infinite 1-D PE chain as a function of temperature, along with the absolute $\kappa$ contributions from the four acoustic phonon modes and the $\kappa$ values from the SMRT model; $\kappa$ contributions from optical phonons are negligible and thus not shown. The convergence test at 300 K with increasing mesh size is also shown in the inset. (b) Variation of the axial thermal conductivity of 1-D PE chain at 300 K as a function of chain length $l$. Inset of (b), normalized cumulative axial thermal conductivity with the distribution of phonon mean free path $\Lambda$.

The convergence of $\kappa$ with chain length is shown in Fig. 6(b) and $\kappa$ reaches a plateau at 100 μm and such large convergence length explains lack of reaching it by classical MD simulations[25] (the required number of atoms is as high as 2.4 M). At present, the maximum chain length in the MD simulation is around 10 μm with $\kappa$ still increasing with the chain length. The slow convergence of $\kappa$ of 1-D PE with chain length is elucidated in the inset of Fig. 6(b). The half-contributing mean free path ($\Lambda_{1/2}$) extends from 0.12 μm in bulk PE crystal to 2.2 μm in the 1-D PE. The results indicate that the $\kappa$ of 1-D PE is even more sensitive to the sample size than bulk PE in both MD simulations and experimental measurements.

In order to shed light on the large-finite $\kappa$ of PE chain, the cumulative value with frequency is plotted in Fig. 7(a), and the plateau is reached 16 THz, indicating the acoustic phonons dominate thermal transport. Phonons with frequency lower than 5.5 THz (the cutoff frequency of TA1) contribute only 14% to $\kappa$, which is contrary to the dominant $\kappa$ contribution from the flexural mode (ZA) in graphene (as high as 88%[38]). The effective spectral lifetime [defined in Eq. (9)] in Fig. 7(b) shows $\tau_{eff} \sim \omega^{-1}$ for both TA modes and a constant lifetime for the LA and TWA modes. While the lifetime increases with frequency decrease, the group velocity of TA modes vanishes as $v \sim \omega^{0.5}$ at low wavevectors, as shown in Fig. 7(c). Given the dispersion and these lifetime behaviors, $\kappa$ contribution from the low-frequency TA modes is $\kappa_{TA} \propto \int \omega^{-0.5} d\omega$. Similarly, $\kappa$ contribution from long-wavelength LA and TWA modes is expressed as $\kappa_{LA(TWA)} \propto \int \omega^{0.5} d\omega$. Therefore, both linear and quadratic phonon modes contribute finitely to $\kappa$ and thus lead to a convergent thermal conductivity. The convergence behaviors of TAs and TWA modes are consistent with previous studies on the "rotator model"[31,80], for which the potential energy is proportional to the chain distortion angles. As for the LA mode, the 1-D Fermi-Pasta-Ulam (FPU)[33] are often adopted in the literature to simulate the heat transport and divergence is sometimes observed. However, different from the 1-D FPU model with either 1-D vibrations[81,82] or nearest-neighbor interaction[24], the single-chain lattice vibrates in 3-D space and thus the transverse and twisting motion arise. The additional inter-mode scatterings significantly increases the scattering phase space and serve as dissipative sources of vibrational energy transport.

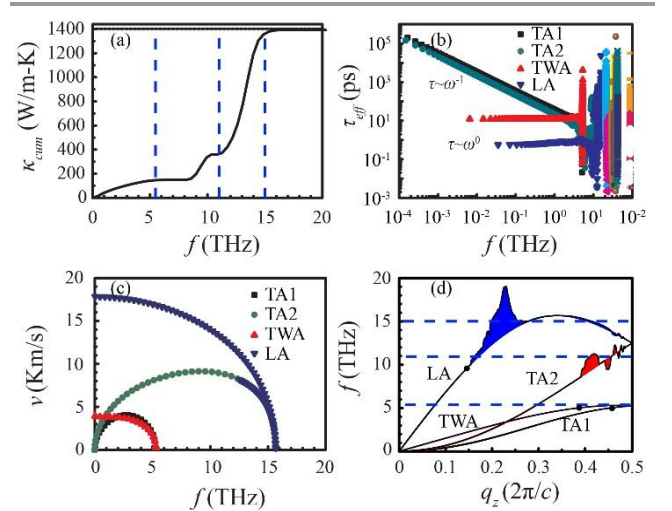

Figure 7. (a) Cumulative axial thermal conductivity of PE chain with respect to frequency. (b) Variations of modal effective lifetime with frequency (logarithmic scales). (c) Variation of the modal group velocity (absolute value) with frequency over 0 to 20 THz. (d) Dispersion with normalized modal contribution to thermal conductivity shown as the vertical broadening of each phonon branch; The three dots represent the three phonons participating in an Umklapp scattering. The three horizontal dashed lines located at 5.5, 11 and 15 THz are used to differentiate phonons and their $\kappa$ contributions. All properties are extracted at the room temperature.

Moreover, from the relative $\kappa$ contribution $\kappa_\lambda$ ($\sum_\lambda \kappa_\lambda = \kappa$) from phonons on each dispersion band in Fig. 7(d) one easily finds that the major room-temperature thermal conductivity is contributed by the LA from 11 to 15 THz, also found in Fig. 6(a) from its dominant contribution. Over this range below the LA cutoff frequency, the LA group velocity maintains relatively large. However, LA modes with lower frequency (5.5 to 11 THz) but larger group velocity only account for a negligible portion of thermal conductivity. This phenomenon is a result of strong Umklapp scattering (total momentum not conserved) LA→TWA+TA1, with the three involved phonons shown as the three dots on Fig. 7(d). This Umklapp scattering channel is blocked once the LA phonon frequency surpasses double of the cutoff of TWA and TA1 and make the phonons above 11 THz overwhelm in $\kappa$ contribution. As for phonons lower than 5.5 THz, even though these long-wavelength phonons hardly experience Umklapp scatterings, there exist strong normal (momentum-conserving) scatterings, for example the intrinsically satisfied collinear scattering LA→LA+LA, TWA→TWA+TWA and other noncollinear scatterings (from Eq (13), normal collinear scattering rates are close to infinite). The strong normal scattering rate can also be observed from the large gap between $\kappa$ from the SMRT model and that from the full solution of BTE in Fig. 6(a). The large normal scatterings contribute indirectly to energy dissipation by dragging phonons to the frequency range where Umklapp scattering is more probable, and thus finally lead to a small effective relaxation time for TWA and LA, as in Fig. 7(b). Moreover, such large normal scatterings may have secondary influence in heat transport under conditions away from static equilibrium (e.g. in MD simulations and thermal transport with temperature gradient) since they may give rise to second sound[79], which implies undamped heat pulse transport, and might lead to a divergent thermal conductivity.

## CONCLUSIONS

We calculated the thermal conductivity of bulk PE crystal and 1-D PE chain by solving the Boltzmann transport equation with the first-principles interatomic force constants. The axial thermal conductivity of bulk PE crystal is 237 W/m-K at room temperature and reduces by 80% in the synthesized nanocrystallites, implying potential for high PE thermal conductivity in high-quality crystallization. Free from the strong anharmonic inter-chain van der Waals forces, a single PE chain has much higher thermal conductivity (1400 W/m-K). Although high, this thermal conductivity is finite, due to the cross-couplings between different phonon branches. The dominant thermal conductivity contribution is from the high-frequency LA phonons, due to the relatively large group velocity and low scattering rate with the low-frequency phonons. The phonon transport analysis guides the nano-design of polymers for enhanced thermal performance.